\documentclass[12pt]{article}
\def\be{\begin{equation}}
\def\ee{\end{equation}}
\def\bea{\begin{eqnarray}}
\def\eea{\end{eqnarray}}
\usepackage{graphicx}

\catcode`\@=11
\def\lsim{\mathrel{\mathpalette\@versim<}}
\def\gsim{\mathrel{\mathpalette\@versim>}}
\def\@versim#1#2{\vcenter{\offinterlineskip
\ialign{$\m@th#1\hfil##\hfil$\crcr#2\crcr\sim\crcr } }}
\catcode`\@=12

\catcode`@=12
\begin{document}
\thispagestyle{empty}
\begin{flushright}
UCRHEP-T520\\
May 2012\
\end{flushright}
\vspace{1.0in}
\begin{center}
{\Large \bf New Simple $A_4$ Neutrino Model for\\ 
Nonzero $\theta_{13}$ and Large $\delta_{CP}$\\}
\vspace{0.5in}
{\bf Hajime Ishimori$^1$ and Ernest Ma$^{2,3}$\\}
\vspace{0.2in}
{\sl $^1$ Department of Physics, Kyoto University,
Kyoto 606-8502, Japan\\}
\vspace{0.1in}
{\sl $^2$ Department of Physics and Astronomy, University of
California,\\
Riverside, California 92521, USA\\}
{\sl $^3$  Kavli Institute for the Physics and Mathematics of the Universe,\\ 
University of Tokyo, Kashiwa 277-8583, Japan\\}
\end{center}
\vspace{01.5in}
\begin{abstract}\
In a new simple application of the non-Abelian discrete symmetry $A_4$ to 
charged-lepton and neutrino mass matrices, we show that for the current 
experimental central value of $\sin^2 2 \theta_{13} \simeq 0.1$, leptonic 
$CP$ violation is necessarily large, i.e. $|\tan \delta_{CP}| > 1.3$.
\end{abstract}

\newpage
\baselineskip 24pt

The non-Abelian discrete symmetry $A_4$ was introduced~\cite{mr01,m02,bmv03} 
to achieve the seemingly impossible, i.e. the existence of a lepton family 
symmetry consistent with the three very different charged-lepton masses 
$m_e$, $m_\mu$, $m_\tau$.  It was subsequently shown~\cite{m04} to be a 
natural theoretical framework for neutrino tribimaximal mixing, i.e. 
$\sin^2 \theta_{23} = 1$, $\tan^2 \theta_{12} = 0.5$, and $\theta_{13}=0$. 
This pattern was consistent with experimental data until recently, when 
the Daya Bay Collaboration reported~\cite{daya12} the first precise 
measurement of $\theta_{13}$, i.e.
\begin{equation}
\sin^2 2 \theta_{13} = 0.092 \pm 0.016({\rm stat}) \pm 0.005({\rm syst}),
\end{equation}
followed shortly~\cite{reno12} by the RENO Collaboration, i.e.
\begin{equation}
\sin^2 2 \theta_{13} = 0.113 \pm 0.013({\rm stat}) \pm 0.019({\rm syst}).
\end{equation}
This means that tribimaximal mixing is not a good description, and more 
importantly, leptonic $CP$ violation is now possible because $\theta_{13} 
\neq 0$, just as hadronic $CP$ violation in the quark sector is possible 
because $V_{ub} \neq 0$.

In this paper, we show that $A_4$ is still a good symmetry for understanding 
this pattern, using a new simple variation of the original idea.  As shown 
below, it predicts a correlation between $\theta_{13}$, $\theta_{23}$, and 
$\delta_{CP}$ in such a way that given the experimentally allowed ranges 
of values for $\theta_{13}$ and $\theta_{23}$, a lower bound on 
$|\tan \delta_{CP}|$ is obtained.  In particular, for the central values 
of $\sin^2 2 \theta_{13} = 0.1$ and $\sin^2 2 \theta_{12} = 0.87$, 
we find $|\tan \delta_{CP}| > 1.3$ from $\sin^2 2 \theta_{23} > 0.92$.

The most general $3 \times 3$ Majorana neutrino mass matrix has six complex
entries, i.e. twelve parameters.  Three are overall phases of the mass
eigenstates which are unobservable.  The nine others are three masses,
three mixing angles, and three phases: one Dirac phase $\delta_{CP}$, i.e.
the analog of the one complex phase of the $3 \times 3$ quark mixing matrix,
and two relative Majorana phases $\alpha_{1,2}$ for two of the three mass
eigenstates. The existence of nonzero $\delta_{CP}$ or $\alpha_{1,2}$ means
that $CP$ conservation is violated.  It is one of the most important issues of
neutrino physics yet to be explored experimentally.

Before showing how the $A_4$ model is constructed, consider first the end 
results.  In the $A_4$ basis, the $3 \times 3$ charged-lepton mass matrix is 
\begin{equation}
{\cal M}_l = {1 \over \sqrt{3}} \pmatrix{1 & 1 & 1 \cr 1 & \omega^2 & \omega 
\cr 1 & \omega & \omega^2} \pmatrix{m_e & 0 & 0 \cr 0 & m_\mu & 0 \cr 
0 & 0 & m_\tau},
\end{equation}
where $\omega = e^{2 \pi i/3} = -1/2 + i \sqrt{3}/2$, and the Majorana 
neutrino mass matrix is
\begin{equation}
{\cal M}_\nu = \pmatrix{a & f & e \cr f & a & d \cr e & d & a}.
\end{equation}
Consider now the tribimaximal basis, i.e.
\begin{equation}
\pmatrix{\nu_e \cr \nu_\mu \cr \nu_\tau} = \pmatrix{\sqrt{2/3} & 1/\sqrt{3} 
& 0 \cr -1/\sqrt{6} & 1/\sqrt{3} & -1/\sqrt{2} \cr -1/\sqrt{6} & 1/\sqrt{3} 
& 1/\sqrt{2}} \pmatrix{\nu_1 \cr \nu_2 \cr \nu_3},
\end{equation}
then
\begin{equation}
{\cal M}_\nu^{(1,2,3)} = \pmatrix{a+d & b & 0 \cr b & a & c \cr 0 & c & a-d},
\end{equation}
where $b=(e+f)/\sqrt{2}$, $c=(e-f)/\sqrt{2}$.  The advantage of using this 
basis is that the experimental values of the mixing angles are not too far 
from the tribimaximal pattern, so that the unitary matrix which diagonalizes 
${\cal M}_\nu^{(1,2,3)}$ may be approximated by
\begin{equation}
U_\epsilon \simeq \pmatrix{1 & \epsilon_{12} & \epsilon_{13} \cr 
-\epsilon_{12}^* 
& 1 & \epsilon_{23} \cr -\epsilon_{13}^* & -\epsilon_{23}^* & 1}.
\end{equation}
Suppose the parameters $a,b,c,d$ are all real in Eq.~(6), then for small 
$b,c$, we find
\begin{equation}
\epsilon_{12} \simeq {b \over d}, ~~~ \epsilon_{23} \simeq {c \over d}, 
~~~ \epsilon_{13} \simeq 0.
\end{equation}
This implies
\begin{equation}
\tan^2 \theta_{12} \simeq (1 - 3 \sqrt{2} \epsilon_{12})/2, ~~~ 
\sin^2 2 \theta_{23} \simeq 1 - (8 \epsilon_{23}^2/3), ~~~ 
\sin \theta_{13} \simeq -\epsilon_{23}/\sqrt{3}.
\end{equation}
We then have the prediction
\begin{equation}
\sin^2 2 \theta_{23} \simeq 1 - 2 \sin^2 2 \theta_{13}.
\end{equation}
Using the existing bound~\cite{pdg10} of $\sin^2 2 \theta_{23} > 0.92$, 
this would 
require $\sin^2 2 \theta_{13} < 0.04$, which is of course ruled out 
by the recent data, i.e Eqs.~(1) and (2).  This result is however not 
negative, but rather very positive, because it says that $\epsilon_{23}$ 
must be complex, in which case the approximation becomes 
\begin{equation}
\sin^2 2 \theta_{23} \simeq 1 - 8 [Re(U_{e3})]^2.
\end{equation}
Now the new data can be accommodated provided that leptonic 
$CP$ violation is large.

In analyzing Eq.~(6), we note from Eq.~(4) that whereas the parameter $a$ 
may be chosen real, the others $b,c,d$ must be kept complex.  In fact, 
even in the tribimaximal limit $(b=c=0)$, $d$ is in general complex, as 
shown already some time ago~\cite{m05}.

We now show how Eqs.~(3) and (4) are obtained.  The symmetry $A_4$ is that of 
the even permutation of four objects.  It has twelve elements and is the 
smallest group which admits an irreducible three-dimensional representation. 
Its character table is given below.
\begin{table}[htb]
\begin{center}
\begin{tabular}{|c|c|c|c|c|c|c|}
\hline
class & $n$ & $h$ & $\chi_1$ & $\chi_{1'}$ & $\chi_{1''}$ & $\chi_3$ \\
\hline
$C_1$ & 1 & 1 & 1 & 1 & 1 & 3 \\
$C_2$ & 4 & 3 & 1 & $\omega$ & $\omega^2$ & 0 \\
$C_3$ & 4 & 3 & 1 & $\omega^2$ & $\omega$ & 0 \\
$C_4$ & 3 & 2 & 1 & 0 & 0 & --1 \\
\hline
\end{tabular}
\caption{Character table of $A_4$.}
\end{center}
\end{table}
The basic multiplication rule of $A_4$ is
\begin{equation}
\underline{3} \times \underline{3} = \underline{1} + \underline{1}' + 
\underline{1}'' + \underline{3} + \underline{3}.
\end{equation}
As first shown in Ref.~\cite{mr01}, for $(\nu_i,l_i) \sim \underline{3}$, 
$l^c_i \sim \underline{1}, \underline{1}', \underline{1}''$, and 
$\Phi_i = (\phi^0_i, \phi^-_i) \sim \underline{3}$, the charged-lepton 
mass matrix is given by
\begin{equation}
{\cal M}_l = \pmatrix{v_1 & 0 & 0 \cr 0 & v_2 & 0 \cr 0 & 0 & v_3} 
\pmatrix{1 & 1 & 1 \cr 1 & \omega^2 & \omega \cr 1 & \omega & \omega^2} 
\pmatrix{f_1 & 0 & 0 \cr 0 & f_2 & 0 \cr 0 & 0 & f_3},
\end{equation}
where $v_i = \langle \phi_i^0 \rangle$.  For $v_1=v_2=v_3=v/\sqrt{3}$, 
we then obtain Eq.(3) with $m_e = f_1 v$, $m_\mu = f_2 v$, $m_\tau = f_3 v$. 
The original $A_4$ symmetry is now broken to the residual symmetry $Z_3$, 
i.e. lepton flavor triality~\cite{m10}, with $e \sim 1$, $\mu \sim \omega^2$, 
$\tau \sim \omega$.  This is a good symmetry of the Lagrangian as long as 
neutrino masses are zero.  Exotic scalar decays are predicted and may be 
observable at the Large Hadron Collider (LHC) in some regions of parameter 
space~\cite{ckmo11,cdmw11}.

To obtain nonzero neutrino masses, we add four Higgs triplets: 
$\Delta_0 \sim \underline{1}$, $\Delta_i \sim \underline{3}$ under $A_4$. 
Let $\langle \Delta^0_0 \rangle = u_0$,  $\langle \Delta^0_1 \rangle = u_i$,  
then Eq.~(4) is the automatic result.  In previous studies, $e=f=0$ has to 
be enforced to get tribimaximal mixing, which is technically an unnatural 
condition, requiring usually the addition of extra symmetries and auxiliary 
fields.  Free of this burden, nonzero and arbitrary $d,e,f$ are easily 
implemented.  For large Higgs triplet masses, small vacuum expectation 
values are naturally induced~\cite{ms98} by the soft trilinear 
$\tilde{\Phi}^\dagger \Delta \Phi$ terms.  We simply assume that $A_4$ is 
broken completely by these terms to obtain different $u_{1,2,3}$.  On the 
other hand, the tribimaximal requirement of $u_2=u_3=0$ is very difficult 
to maintain, because it is not protected against infinite radiative 
corrections, which is the field theory's way of telling us that they 
should be nonzero and arbitrary in the first place.  In retrospect, it 
should have been obvious that Eq.~(4) is the more natural choice for the 
neutrino mass matrix in the $A_4$ basis.

The most general neutrino mass matrix in the tribimaximal basis is 
\begin{equation}
{\cal M}_\nu^{(1,2,3)} = \pmatrix{m_1 & m_6 & m_4 \cr m_6 & m_2 & m_5 
\cr m_4 & m_5 & m_3}.
\end{equation}
To first order, $\theta_{13}$ and $\theta_{23}$ are sensitive to $m_4$ and 
$m_5$, whereas $\theta_{12}$ is sensitive to $m_6$.  The case 
of $m_6=0$ was considered in the original proposal~\cite{m04} of 
tribimaximal mixing using $A_4$, and updated recently~\cite{mw11}. 
The case of $m_5 \simeq 0$ is realized in a supersymmetric $B-L$ gauge 
model with $T_7$ symmetry discussed recently~\cite{ckmo11R,ikm12}.  
The case of unbroken residual symmetries in a class of discrete symmetries 
has been discussed recently~\cite{hs12}, as well as a general perturbation 
of the tribimaximal limit~\cite{br12}.  
Here we consider the simplest and perhaps the most compelling case of 
$m_4=0$, which does not correspond to any unbroken residual symmetry. 
The fact that $m_4=0$ is simply the result of not having Higgs triplets 
which transform as $\underline{1}'$ or $\underline{1}''$ under $A_4$.

The neutrino mixing matrix $U$ has 4 parameters: $s_{12}, s_{23}, s_{13}$ and
$\delta_{CP}$~\cite{pdg10}.  We choose the convention $U_{\tau 1}, U_{\tau 2},
U_{e3}, U_{\mu 3} \to -U_{\tau 1}, -U_{\tau 2},
-U_{e3}, -U_{\mu 3}$ to conform with that of the tribimaximal mixing matrix
of Eq.~(5), then
\begin{equation}
{\cal M}_\nu^{(1,2,3)} = U^T_{TB} U \pmatrix{e^{i\alpha_1} m'_1 & 0 & 0 \cr
0 & e^{i\alpha_2} m'_2 & 0 \cr 0 & 0 &  m'_3} U^T U_{TB},
\end{equation}
where $m'_{1,2,3}$ are the physical neutrino masses, with
\begin{eqnarray}
m'_2 &=& \sqrt{{m'_1}^2 + \Delta m^2_{21}}, \\
m'_3 &=& \sqrt{{m'_1}^2 + \Delta m^2_{21}/2 + \Delta m_{32}^2}~~{\rm (normal
~hierarchy)}, \\
m'_3 &=& \sqrt{{m'_1}^2 + \Delta m^2_{21}/2 - \Delta m_{32}^2}~~{\rm
(inverted~hierarchy)}.
\end{eqnarray}
If $U$ and $\alpha_{1,2}$ are known, then all $m_{1,2,3,4,5,6}$ are functions
only of $m'_1$.

We now diagonalize ${\cal M}_\nu^{(1,2,3)}$ using
\begin{equation}
U_\epsilon {\cal M}_\nu^{(1,2,3)} U_\epsilon^T = \pmatrix{e^{i \alpha'_1}
m'_1 & 0 & 0 \cr 0 & e^{i \alpha'_2} m'_2 & 0 \cr 0 & 0 & e^{i \alpha'_3} 
m'_3},
\end{equation}
from which we obtain $U' = U_{TB} U^T_\epsilon$.  To obtain $U$ with the 
usual convention, we rotate the phases of the $\mu$ and $\tau$ rows so that
$U'_{\mu 3} e^{-i \alpha'_3/2}$ is real and negative, and 
$U'_{\tau 3} e^{-i \alpha_3/2}$ is real and positive.
These phases are absorbed by the $\mu$ and $\tau$ leptons and are
unobservable.  We then rotate the $\nu_{1,2}$ columns so that 
$U'_{e1} e^{-i \alpha_3/2} = U_{e1} e^{i \alpha''_1/2}$ and 
$U'_{e2} e^{-i \alpha_3/2} = U_{e2} e^{i \alpha''_2/2}$, where 
$U_{e1}$ and $U_{e2}$ are real and positive.  
The physical relative Majorana phases of $\nu_{1,2}$ are then 
$\alpha_{1,2} = \alpha'_{1,2} + \alpha''_{1,2}$.  The three angles and 
the Dirac phase are extracted according to
\begin{equation}
\tan^2 \theta_{12} = |U'_{e1}/U'_{e2}|^2, ~~~ 
\tan^2 \theta_{23} = |U'_{\mu 3}/U'_{\tau 3}|^2, ~~~ 
\sin \theta_{13} e^{-i\delta_{CP}} = U'_{e3} e^{-i \alpha'_3/2}.
\end{equation}
The effective Majorana neutrino mass in neutrinoless double beta decay is 
then given by
\begin{equation}
m_{ee} = |U_{e1}^2 e^{i \alpha_1} m'_1 + U_{e2}^2 e^{i \alpha_2}
m'_2 + U_{e3}^2 m'_3|.
\end{equation}

Although $b,c,d$ are in general complex, the structure of this model is 
restricted by data such that $Im(b)$ is very small, so we will assume in the 
following that $b$ is real.  As for $Im(d)$, it is also small and affects 
only $m_{ee}$ slightly and not $\delta_{CP}$, so we will also take $d$ 
to be real.  The main feature here is the complexity of $c$.  To first 
approximation, we find
\begin{equation}
d \simeq -a, ~~~ U_{e3} \simeq -{Re(c) \over 2a} + i {Im(c) \over 4a},
\end{equation}
allowing only the normal ordering of neutrino masses.

The special case $b=0$ is especially interesting.  It may be maintained 
by an interchange symmetry~\cite{m04,mw11} such that $f = -e$.  As such, 
it was considered in Ref.~\cite{hs12}. 
In that case, Eq.~(6) can be diagonalized exactly. 
Assuming that $a,d$ are real and $c$ 
complex, we find
\begin{eqnarray}
\tan^2 \theta_{12} &=& {1 - 3 \sin^2 \theta_{13} \over 2}, \\ 
\tan^2 \theta_{23} &=& { \left(1- {\sqrt{2} \sin \theta_{13} \cos \delta'_{CP} 
\over \sqrt{1-3 \sin^2 \theta_{13}}} \right)^2 + {2 \sin^2 \theta_{13} 
\sin^2 \delta'_{CP} \over 1-3 \sin^2 \theta_{13}} \over 
\left(1+ {\sqrt{2} \sin \theta_{13} \cos \delta'_{CP} 
\over \sqrt{1-3 \sin^2 \theta_{13}}} \right)^2 + {2 \sin^2 \theta_{13} 
\sin^2 \delta'_{CP} \over 1-3 \sin^2 \theta_{13}}},
\end{eqnarray}
where $\delta'_{CP} = \delta_{CP} - \alpha'_3/2$.  The phase $\alpha'_3$ 
is defined in Eq.~(19) and depends on the specific values of Eq.~(6). 
For $\sin \theta_{13} = 0.16$, corresponding to $\sin^2 2 \theta_{13} = 0.1$, 
this predicts $\tan^2 \theta_{12} = 0.46$.   If $\delta_{CP}=0$ (which also 
implies that $\alpha'_3 = 0$), then this 
would also predict $\sin^2 2 \theta_{23} = 0.80$ which is of course ruled 
out.  Using $\sin^2 2 \theta_{23} > 0.92$, we find in this case 
$|\tan \delta'_{CP}| > 1.2$.

For our numerical analysis, we set
\begin{eqnarray}
&& \Delta m^2_{21} = 7.59 \times 10^{-5}~{\rm eV}^2, ~~~ 
\Delta m^2_{32} = 2.45 \times 10^{-3}~{\rm eV}^2, \\ 
&& \sin^2 2 \theta_{12} = 0.87, ~~~  
\sin^2 2 \theta_{13} = 0.05~{\rm to}~0.15.  
\end{eqnarray}
We then diagonalize Eq.~(6) exactly and scan for solutions satisfying the 
above experimental inputs.   We do not assume $b=0$ or $\alpha'_3$ to 
be necessarily small.  We find that solutions exist only for the 
normal ordering of neutrino masses, i.e. $m'_1 < m'_2 < m'_3$, as in 
the tribimaximal case~\cite{m05}.  In Fig.~1 we show $|\tan \delta_{CP}|$ 
as a function of $\sin^2 2 \theta_{13}$ from 0.05 to 0.15, for the central 
value of $\sin^2 2 \theta_{12} = 0.87$ and the two fixed values of 
$\sin^2 2 \theta_{23} = 0.92$ and 0.96.  In Fig.~2 we plot the parameter $b$ 
as a function of $\sin^2 2 \theta_{13}$.  It shows that for 
$\sin^2 2 \theta_{12} = 0.87$, it is indeed very small.  Note that for 
$b=0$, we find $\sin^2 2 \theta_{13} = 0.08$, i.e. $\sin^2 \theta_{13} = 
0.02$.  Using Eq.~(23), we recover exactly $\tan^2 \theta_{12} = 0.47$, 
i.e. $\sin^2 2 \theta_{12} = 0.87$, as expected.  In Figs.~3 and 4 we plot 
the physical neutrino masses $m'_{1,2,3}$ together with the effective 
neutrino mass $m_{ee}$ in neutrinoless double beta decay as functions 
of $\sin^2 2 \theta_{13}$ for $\sin^2 2 \theta_{23} = 0.92$ and 0.96 
respectively.  Note that $m_{ee}$
is always smaller than $m'_1$ because $e^{i \alpha_1} = -1$ and 
$e^{i \alpha_2} = 1$ in Eq.~(21).  In Figs.~5 and 6 we plot the parameters 
$a,-d,Re(c),Im(c)$ as functions of 
of $\sin^2 2 \theta_{13}$ for $\sin^2 2 \theta_{23} = 0.92$ and 0.96 
respectively.

In conclusion, neutrino tribimaximal mixing may be dead, but $A_4$ is 
alive and even getting healthier.  In a new simple application, given the 
present allowed ranges of values for $\sin^2 2 \theta_{13}$ and 
$\sin^2 2 \theta_{23}$, we predict large $CP$ violation and a normal 
ordering of neutrino masses.

\underline{Acknowledgments}: 
The work of H.I. is supported by Grant-in-Aid for Scientific Research, 
No.~23.696, from the Japan Society of Promotion of Science.
The work of E.M. is supported in part by the U.~S.~Department of Energy 
under Grant No.~DE-AC02-06CH11357.

\newpage
\bibliographystyle{unsrt}

\begin{figure}[htb]
\includegraphics[width=14cm]{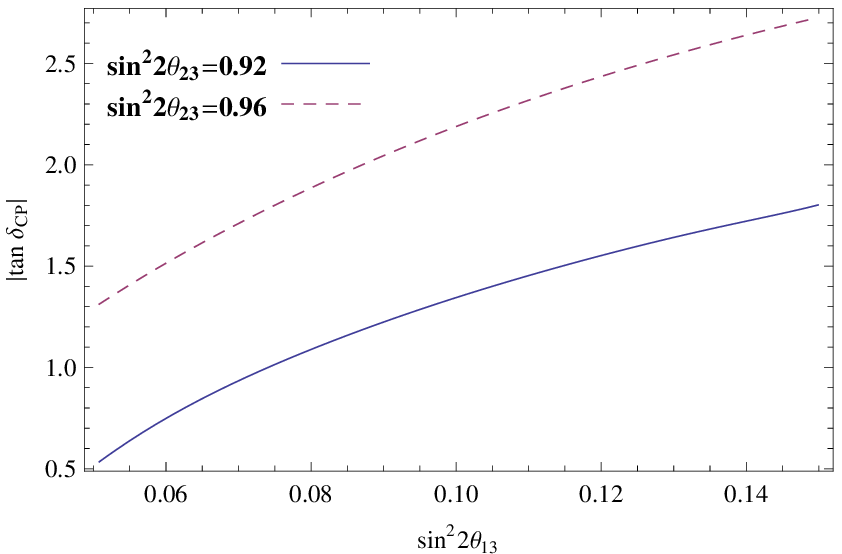}
\caption{$|\tan \delta_{CP}|$ versus $\sin^2 2 \theta_{13}$ for 
$\sin^22\theta_{23}=0.92$ and $0.96$.}
\end{figure}
\begin{figure}[htb]
\includegraphics[width=14cm]{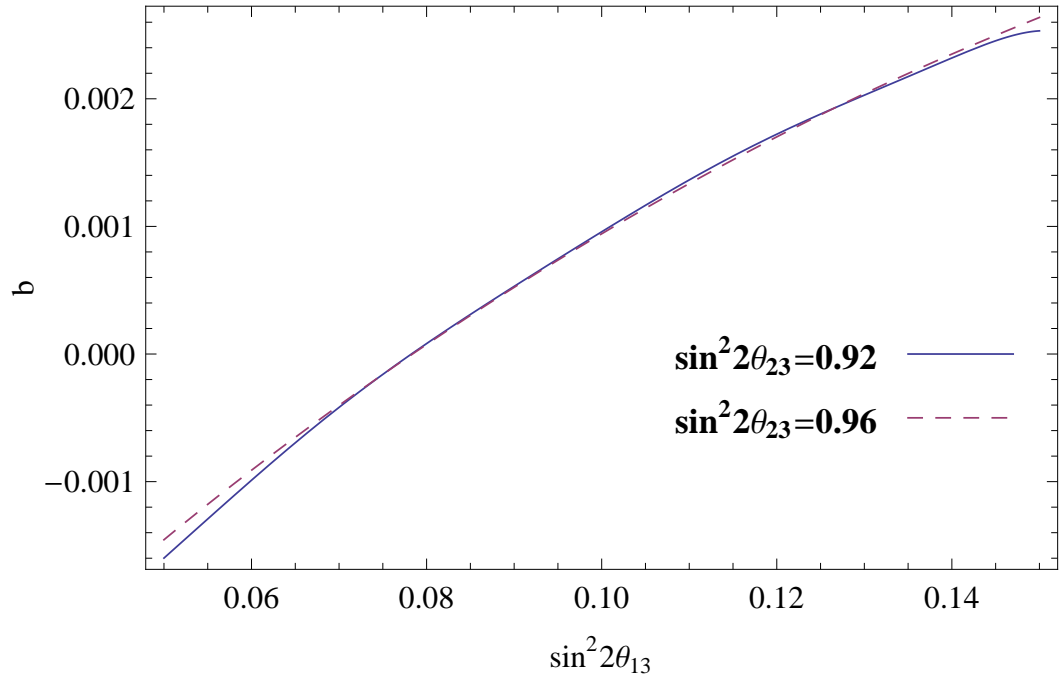}
\caption{Parameter $b$ versus $\sin^22\theta_{13}$ for  
$\sin^22\theta_{23}=0.92$ and $0.96$.}
\end{figure}
\begin{figure}[htb]
\includegraphics[width=14cm]{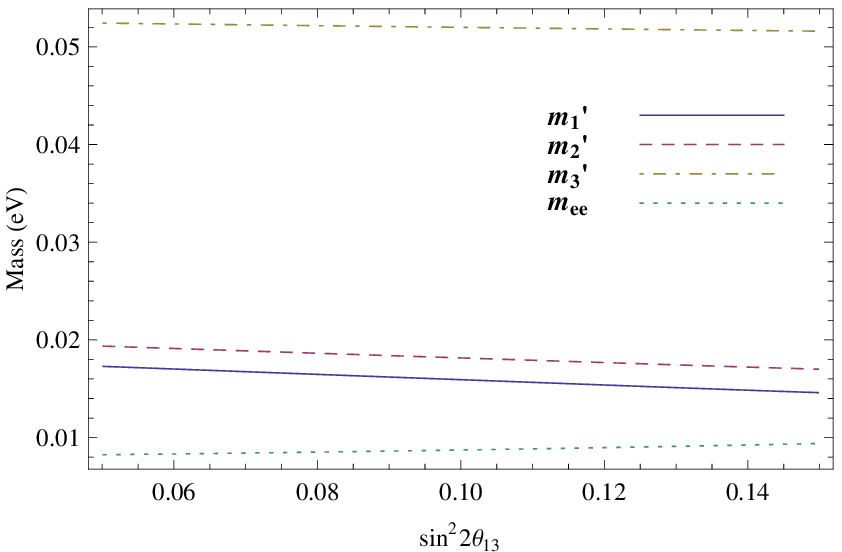}
\caption{Physical neutrino masses and the effective neutrino mass $m_{ee}$ 
in neutrinoless double beta decay for  
$\sin^22\theta_{23}=0.92$.}
\end{figure}
\begin{figure}[htb]
\includegraphics[width=14cm]{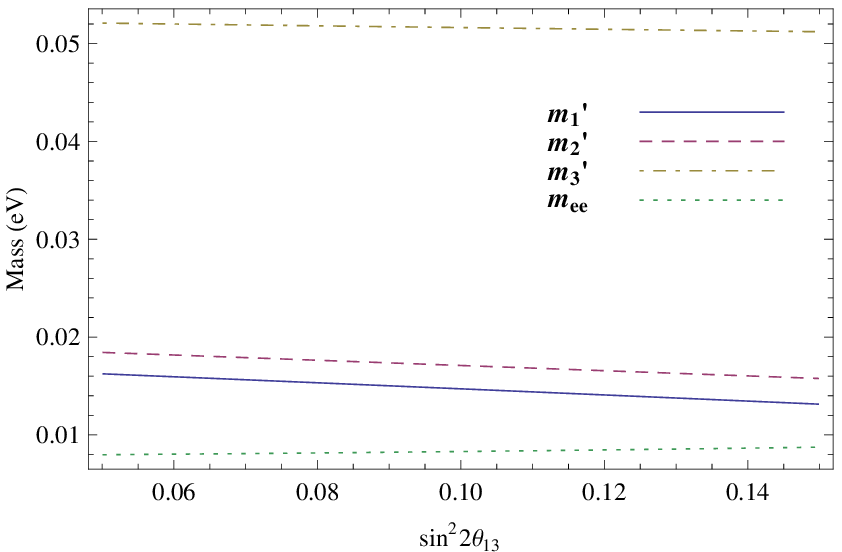}
\caption{Physical neutrino masses and the effective neutrino mass $m_{ee}$ 
in neutrinoless double beta decay for 
$\sin^22\theta_{23}=0.96$.}
\end{figure}
\begin{figure}[htb]
\includegraphics[width=14cm]{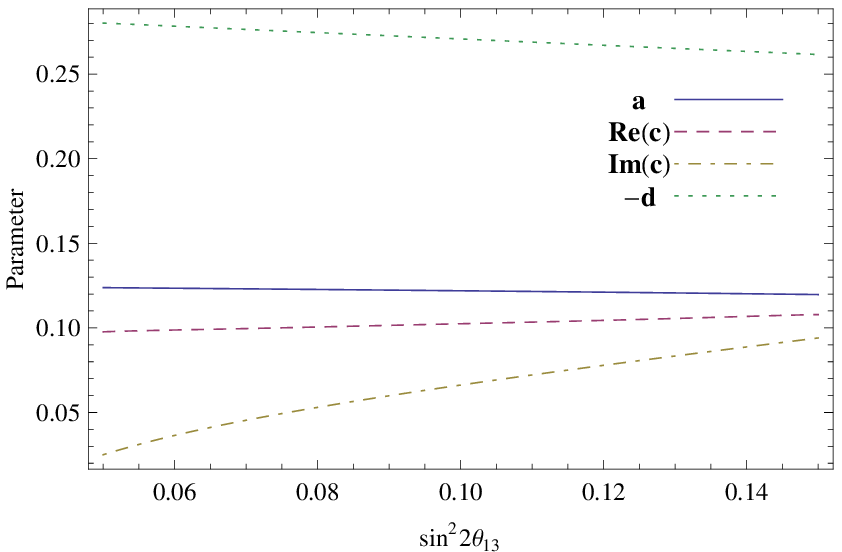}
\caption{$A_4$ parameters for 
$\sin^22\theta_{23}=0.92$.}
\end{figure}
\begin{figure}[htb]
\includegraphics[width=14cm]{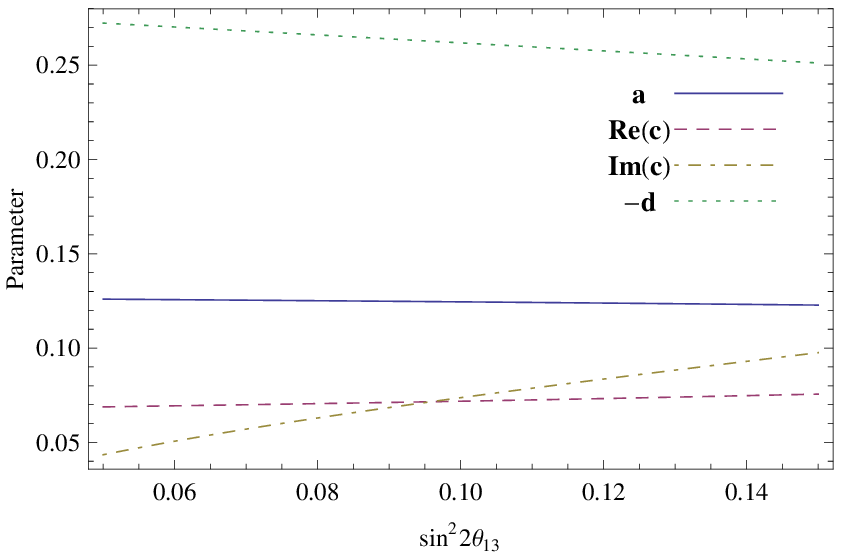}
\caption{$A_4$ parameters for  
$\sin^22\theta_{23}=0.96$.}
\end{figure}

\end{document}